\documentclass[preprint,12pt]{elsarticle}
\usepackage{multirow} 
\usepackage{booktabs} 
\usepackage{amssymb}
\usepackage{amsmath}
\journal{Medical Physics \& Engineering}
\usepackage{setspace}
\usepackage{graphicx}
\usepackage{amsmath}
\usepackage{color}

\begin{document}

\begin{frontmatter}

\title{Is there any information on micro-structure in microwave tomography of bone tissue?}

\author[IFLY,GAMEFI]{R.M.Irastorza}
\author[IFLY,UTN]{C. M. Carlevaro}
\author[IFLY,GAMEFI]{F. Vericat}
\address[IFLY]{Instituto de F\'{i}sica de L\'{i}quidos y Sistemas Biol\'{o}gicos (IFLYSIB-CCT La Plata-CONICET), Calle 59 No 789, B1900BTE La Plata, Argentina.}
\address[UTN]{Universidad Tecnol\'{o}gica Nacional - FRBA, UDB F\'{i}sica, Mozart No 2300, C1407IVT Buenos Aires, Agentina.}
\address[GAMEFI]{Grupo de Aplicaciones Matem\'aticas y Estad\'isticas de la Facultad de Ingenier\'ia Universidad Nacional de La Plata, Argentina.}

\begin{abstract}

In this work, two-dimensional simulations of the microwave dielectric properties of models with ellipses and realistic models of trabecular bone tissue are performed. In these simulations, finite difference time domain methodology has been applied to simulate two-phase structures containing inclusions. The results presented here show that the micro-structure is an important factor in the effective dielectric properties of trabecular bone. We consider the feasibility of using the dielectric behaviour of bone tissue to be an indicator of bone health. The frequency used was 950  MHz. It was found that the dielectric properties can be used as an estimate of the degree of anisotropy of the micro-structure of the trabecular tissue. Conductivity appears to be the most sensitive parameter in this respect. Models with ellipse shaped
-inclusions are also tested to study their application to modelling bone tissue. Models with ellipses that had an aspect ratio of a/b = 1.5 showed relatively good agreement when compared with realistic models of 
bone tissue. According to the results presented here, the anisotropy of trabecular bone must be accounted for when measuring its dielectric properties using microwave imaging.

\end{abstract}


\end{frontmatter}

\section{Introduction}
\label{INTRO}
Osteoporosis is usually diagnosed by one parameter, the bone mineral density (BMD). Unfortunately, this parameter does not provide direct tissue bio-mechanical information. Currently, researchers are searching for alternative methods for the \textit{in vivo} estimation of bone mechanical properties. Ultrasound appears to be the most advanced tool \cite{haidekker2011medical}. Other approaches, such as that developed by Kanis and co-workers \cite{kanis2008assessment}, intend to integrate risks that are associated with clinical risk factors and the BMD at the femoral neck.\\ 
The measurement of dielectric properties of bones has recently received attention because of its potential application as an alternative diagnostic method of bone health \cite{cherkaev2011characterization,bonifasi2009electrical,sierpowska2007effect,sierpowska2007electrical,sierpowska2005prediction,sierpowska2003electrical}. Low frequency \textit{in vitro} measurements on a specific sample (human cadavers, age = 58 $\pm$ 20 years) have shown that the linear combination of two relative permittivities in the frequency ranges (50 Hz-1 kHz and 100 kHz-5  MHz), can predict 73\% of the trabecular bone volume fraction (the ratio of bone volume to total tissue volume BV/TV) \cite{sierpowska2006interrelationships}. Currently, low frequency methods are applicable only to special cases of open surgery. Recently, the first relative permittivity and conductivity of \textit{in vivo} images of 
human calcaneus have been obtained by microwave tomography \cite{golnabi2011microwave}. Furthermore, \textit{in vitro} 
microwave studies suggest a negative correlation between BMD and permittivity / conductivity \cite{meaney2012bone,irastorzatesis,peyman2011dielectric}. In reference \cite{peyman2011dielectric}, it is hypothesised that mineralised bones have a lower relative permittivity mainly because they have lower water content.\\
It is well known that not only mineral content but also micro-structure defines the mechanical and dielectric properties of bone tissue \cite{sierpowska2006interrelationships}. There are several parameters for characterising the micro-structure of bone, including the following: BV / TV, the mean intercept length (MIL) and the trabecular thickness \cite{Odgaard_book}. In reference \cite{cherkaev2011characterization}, a theoretical approach to porosity estimation is developed. The authors use effective dielectric properties to obtain the spectral function of the bone, which is closely related to its porosity.\\
The age and pathological conditions produce selective bone tissue resorption \cite{orwoll1999osteoporosis}, which results in thinner horizontal trabeculae. Efforts have been made to detect this anisotropy by using ultrasound techniques \cite{hosokawa2009numerical,hosokawa2010effect}. In these papers, the author uses fast and slow waves, which can propagate in the direction of the major trabecular orientation. Trabecular bone was found to be electrically anisotropic. The relationships between the dielectric properties at low frequencies ($<$ 10  MHz) in different directions have been studied by Saha et. al \cite{saha1995comparison}. These variations could be interpreted as changes in the trabecular orientation, but there is no strong evidence for this interpretation yet. In contrast, at low frequencies ($<$ 10 MHz), no linear correlation between the degree of anisotropy and the dielectric or electrical parameters was found by Sierpowska et al. \cite{sierpowska2006interrelationships}.\\
In electromagnetic simulation and the theory of hierarchical materials, homogenisation techniques are used extensively \cite{teixeira2008time}. These techniques provide information on micro-structure \cite{karkkainen2000} and anisotropy \cite{mejdoubi2006}. This paper intends to take a step towards a micro-structure dielectric model of trabecular bone tissue. We consider effective dielectric approaches and their application to realistic electromagnetic models based on microscopic images of bone tissue. The main idea is to study the anisotropy of trabecular micro-structure and its relationships with the dielectric effective parameters using simulation techniques. The two dimensional (2D) models used here are numerically solved by the finite difference time domain (FDTD) technique \cite{oskooi2010,taflove2005}.
We attempt to answer the following question: 
If we measure the trabecular bone in the microwave range, is it warranted to consider isotropic effective dielectric properties? The answer can contribute to the development of a tool for the microstructural characterisation of bone tissue via effective models that can account for anisotropy.\\

\section{Methods}
\label{METH}
\begin{figure}%[h]
\centering
\includegraphics[width=0.8\textwidth]{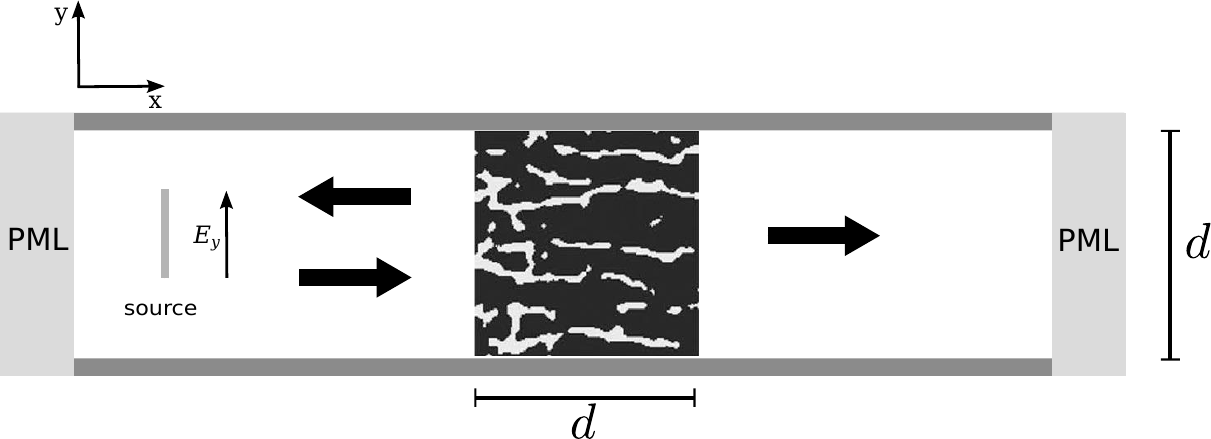}
\caption{Simulation box.}
\label{Simulationbox}
\end{figure}

The effective relative permittivity ($\varepsilon_{\text{eff}}$) calculation in 2D is studied by observing the reflection of a slice of the relative dielectric constant $\varepsilon_{\text{e}}$, with inclusions of the relative permittivity $\varepsilon_{\text{i}}$. The sub-indexes $\text{e}$ and $\text{i}$ denote the environment and inclusion, respectively. Figure \ref{Simulationbox} shows the simulation box.
Two shapes of inclusion are tested: circular and elliptical. For validation purposes, different values of $\varepsilon_{\text{e}}$ ($\sigma_{\text{e}}$) and $\varepsilon_{\text{i}}$ ($\sigma_{\text{i}}$) are tested (where $\sigma_{\text{x}}$ is the conductivity of medium x). In all of the cases, overlaps between inclusions are allowed. For the simulation of realistic models of trabecular bone, we also consider two phases of material: the trabeculae as the inclusions and the bone marrow as the environment. Values of permittivity and conductivity are obtained from \cite{webdielectric}, which is based on \cite{gabriel1996dielectric}.
\subsection{Numerical models}
\label{NumModel}
Maxwell equations are numerically solved using the finite difference time domain (FDTD) technique \cite{taflove2005}. The implementation of the FDTD algorithm is performed using \textsl{meep} (free FDTD simulation software package \cite{oskooi2010}). The slice is a square with variable thickness ($d$). The entire simulation box has a variable length, and it is adjusted  to be long enough to reach three times the thickness of the sample at each side. A temporal Gaussian pulse polarised in the $y$ direction ($E_{y}$) with a central frequency of $950$ MHz is used as the source, which is located far enough from the interface air-slice. The spatial grid resolution of the simulation box is $50$ $\mu$m (or smaller), and the Courant factor is $S=0.5$. In the $x$ direction, the boundary condition is a perfectly matched layer (PML).\\
To obtain the $\varepsilon_{\text{eff}}$, the reflection coefficient ($\Gamma_{\text{i}}(\omega) $) of the medium with inclusions or simulated bone is compared to the reflection coefficient ($\Gamma(\omega,\varepsilon_{\text{eff}}) $) of a homogeneous slice in the frequency range of interest. Minimisation with respect to $\varepsilon_{\text{eff}}$ of the absolute difference ($\arrowvert \Gamma_{\text{i}}(\omega)-\Gamma(\omega,\varepsilon_{\text{eff}}) \arrowvert$) is accomplished using a Newton conjugate-gradient algorithm (implemented in Scipy 0.9.0 \cite{scipy},  an open-source software).

\subsection{Effective permittivity models}
There are several models that represent effective dielectric properties of two-phase materials. \ref{apendix1} shows some examples. In these models, the effective value depends on the fraction that is occupied by the inclusions ($\phi_{\text{i}}$) and on the permittivities (conductivities) $\varepsilon_{\text{e}}$ ($\sigma_{\text{e}}$) and $\varepsilon_{\text{i}}$ ($\sigma_{\text{i}}$). The effective value also depends on the shape of the inclusions and anisotropy. Those models, which account for the shape and anisotropy, are of particular interest for this work.  As discussed by Mejdoubi and Brusseau \cite{mejdoubi2006}, the effective permittivity can be represented by a function $f(\cdot)$:
\begin{equation}
\varepsilon_{\text{eff}}=\varepsilon_{\text{e}}\cdot f\left(\frac{\varepsilon_{\text{i}}}{\varepsilon_{\text{e}}},\phi_{\text{i}},A \right)\approx\varepsilon_{\text{e}}\left(1+\alpha\phi_{\text{i}}+\beta\phi_{\text{i}}^{2}\right)\text{,}
\label{eqMejdoubi}
\end{equation}
where $\alpha$ and $\beta$ depend on the depolarisation factor ($A$) and the permittivity ratio ($\frac{\varepsilon_{\text{i}}}{\varepsilon_{\text{e}}} $). In 2D, $A$ is a functional of the inclusion shape and permittivity ratio only, and $0\leq A \leq 1$. As mentioned above, $\varepsilon_{\text{e}}$ is the environment's relative permittivity and  $\varepsilon_{\text{i}}$ is the relative permittivity of the inclusions. See details in \ref{apendix1}.  Two widely used forms for the function $f(\cdot)$ are the Maxwell Garnett (MG) and the symmetric Bruggeman (SBG) forms (see \cite{mejdoubi2006} and the references therein). For the SBG approach, $\alpha$ and $\beta$ take the following forms:
\begin{equation}
\begin{array}{c}
 \alpha = 1 / [A+1 / (\varepsilon_{\text{i}} / \varepsilon_{\text{e}} -1)]\text{,}\\
\beta_{\text{SBG}} = \frac{\varepsilon_{\text{i}} / \varepsilon_{\text{e}}}{A^{2}\left(\frac{\varepsilon_{\text{i}}}{\varepsilon_{\text{e}}}-1\right)\left(1+\frac{1}{A\left(\frac{\varepsilon_{\text{i}}}{\varepsilon_{\text{e}}}-1\right)} \right)^{3}}\text{.}
\end{array}
\label{alfaybeta}
\end{equation}
From now on, we call the model in Eq.\ref{eqMejdoubi} the M-B (from Mejdoubi and Brusseau). The depolarisation factor intrinsically contains the information on the shape and orientation of the inclusions. For example, for the elliptical inclusions with 90$^{o}$ orientation with respect to the applied electric field, the aspect ratio (the major semi-axis $a$ / minor semi-axis $b$) $a/b = 1/3$ and $\varepsilon_{\text{i}}/\varepsilon_{\text{e}} = 20/2$,  Mejdoubi and Brusseau obtained $A$ equal to 0.732 using $\beta_{\text{MG}}$ or 0.772 using $\beta_{\text{SBG}}$, as defined in \ref{apendix1}. It should be noted that these values were fitted with $\phi_{\text{i}} < 0.1$.
\subsection{Micro-structure Parameters}
\label{MicroParam}
Trabecular bone structure has been studied extensively and there are several parameters to characterize it (see for example reference \cite{Odgaard_book}). In this work, four parameters are studied on 2D images of trabecular bones: BV / TB, the trabecular thickness, the mean intercept length (MIL) and the degree of anisotropy (DA $= \text{MIL}_{\text{max}} / \text{MIL}_{\text{min}}$ ). The former is associated with bone porosity. For the BV / TB and MIL calculation, see \cite{Odgaard_book} and the references therein. The MIL parameter is calculated as a function of an angle $\eta$, where $0^{o}<\eta <180^{o}$. For the trabecular thickness calculation, we draw $n$ parallel lines that are orthogonal to the main trabecular direction (we define the main trabecular direction as the angle $\eta (\text{MIL}_{\text{max}})$) and \
emph{\color{blue}we} measure the average thickness of the intersections of these lines with trabeculae. 
\subsection{Images}
\label{Images}
\begin{figure}%[h]
\centering
\includegraphics[width=0.9\textwidth]{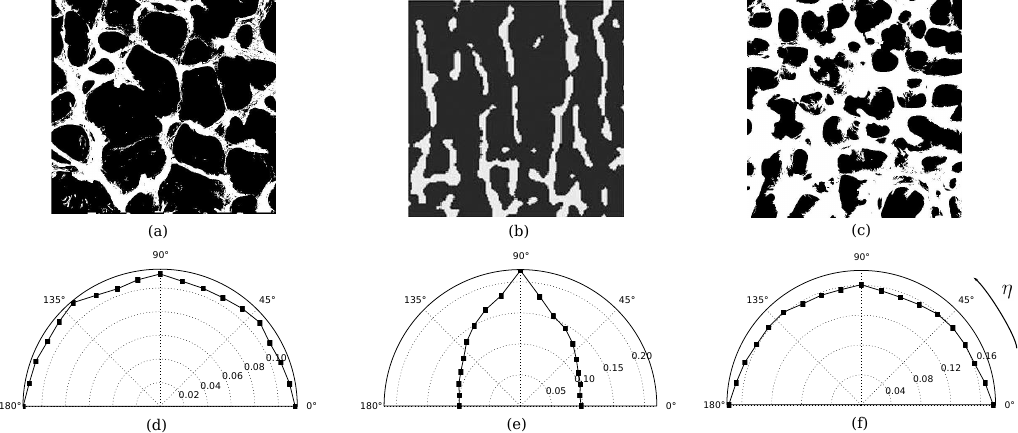}
\caption{Microscopic characterisation of the main direction: the mean intercept length calculation. Parts (a), (b) and (c) correspond to sample 1, 2 and 3 as defined in the text (see subsection \ref{Images}), respectively. (d), (e) and (f) are the polar plots of the MIL parameter and correspond to sample 1, 2 and 3, respectively. The variable $\eta$ is the slope angle of the parallel lines for the MIL calculation. The lines are guides for the eyes.}
\label{MILcalculus}
\end{figure}

The images of trabecular bone tissue used here are available in references \cite{hansmalab,hosokawa2009numerical}. These bitmap images are used to build the realistic dielectric models, which decompose complex shape geometries into triangles (these simple shapes can be simulated with meep \cite{oskooi2010}). Image processing is applied to convert the gray scale images into black and white,  to define a two-phase material using histogram-based segmentation \cite{scikits}.\\
The samples used here are called sample 1, 2, 3, 4, 5 and 6, and are all shown in section \ref{RESUL} (Fig. \ref{samples}). Figure \ref{MILcalculus} shows some of theese samples. Figure \ref{MILcalculus} (a) shows trabecular bone tissue from an old person (sample 1), its porosity is approximately 0.75. The polar plot in Fig. \ref{MILcalculus} (d) shows the corresponding MIL parameter ($\eta$ is the slope angle of the parallel lines for the MIL calculation). The tissue appears to be isotropic; the maximum of the MIL parameter is approximately 0$^{o}$ but the distribution is almost a circle. Figure \ref{MILcalculus} (b) also shows a highly porous old bone, with 0.79 porosity   (sample 2). Clearly, its MIL maximum is approximately 90$^{o}$ (see Fig. \ref{MILcalculus} (e)). Figure \ref{MILcalculus} (c) shows a young bone (sample 3), with a 0.46 porosity and a 0$^{o}$ main direction (Fig. \ref{MILcalculus} (f)).

\section{Results}
\label{RESUL}
\subsection{Validation}
\begin{figure}%[h]
\centering
\includegraphics[width=0.9\textwidth]{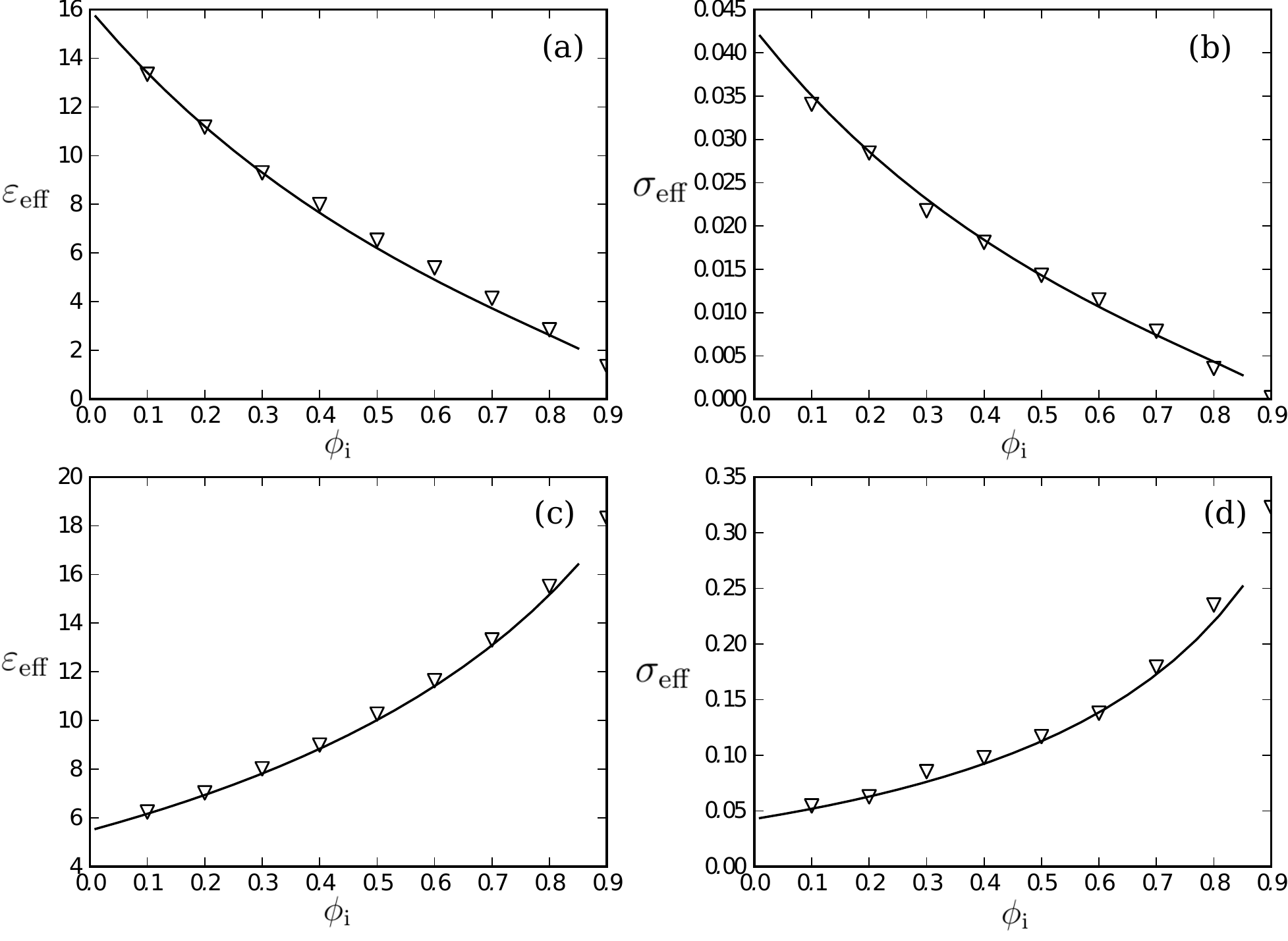}
\caption{The effective permittivity and conductivity of hexagonal array discs as a function of the fraction occupied ($\phi_{\text{i}}$). A comparison between the analytic solution (continuous line) and the simulation ($\triangledown$). Parts (a) and (b) correspond to the effective permittivity and conductivity of a medium with $(\varepsilon_{\text{i}}\text{, }\varepsilon_{\text{e}})=(1.0\text{, }16.0)$ and $(\sigma_{\text{i}}\text{, }\sigma_{\text{e}})=(0.0\text{, }0.043)$. Parts (c) and (d) correspond to the effective permittivity and conductivity of a medium with $(\varepsilon_{\text{i}}\text{, }\varepsilon_{\text{e}})=(20.584\text{, }5.485)$ and $(\sigma_{\text{i}}\text{, }\sigma_{\text{e}})=(0.364\text{, }0.043)$ }
\label{Fighexagonal}
\end{figure}

\begin{figure}%[h]
\centering
\includegraphics[width=0.9\textwidth]{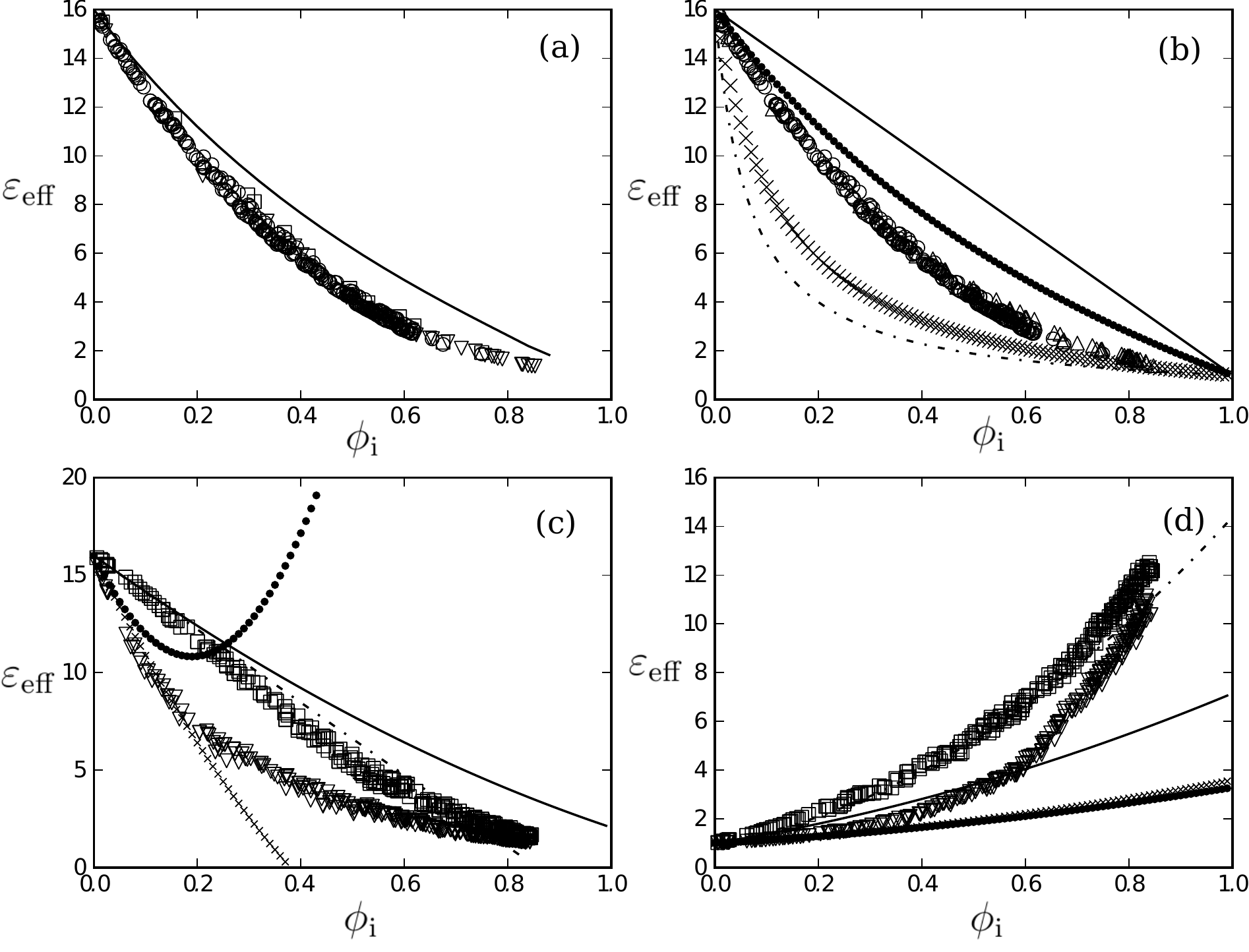}
\caption{Effective relative permittivity of media with disc and ellipse shape inclusions randomly located. (a) Medium with discs for different radius ($r_{1}$, $r_{2}$ and $r_{3}$, $\triangledown$, $\square$ and $\circ$, respectively) compared with analytic solution of hexagonal array (continuous line). $r_{1}<r_{2}<r_{3}$. (b) Effective permittivity of discs ($\circ$) compared with ellipses at $\hat{\gamma}=$ 45$^{o}$ ($\vartriangle$) and Wiener (continuous and dash-dotted lines) and H-S (cross and dotted lines) bounds. (c)-(d) Effective permittivity of oriented ellipses at $\hat{\gamma}=$0$^{o}$ ($\triangledown$), 90$^{o}$ ($\square$) compared with M-B bounds (dotted and continuous marks for $\beta_{\text{MG}}$, cross and dashed-dotted marks for $\beta_{\text{SBG}}$). In (a), (b) and (c) $(\varepsilon_{\text{i}},\varepsilon_{\text{e}})=(1\text{, }16)$. In (d) $(\varepsilon_{\text{i}}\text{, }\varepsilon_{\text{e}})=(16\text{, }1)$.}
\label{elipsescotas}
\end{figure}

\begin{figure}%[h]
\centering
\includegraphics[width=0.9\textwidth]{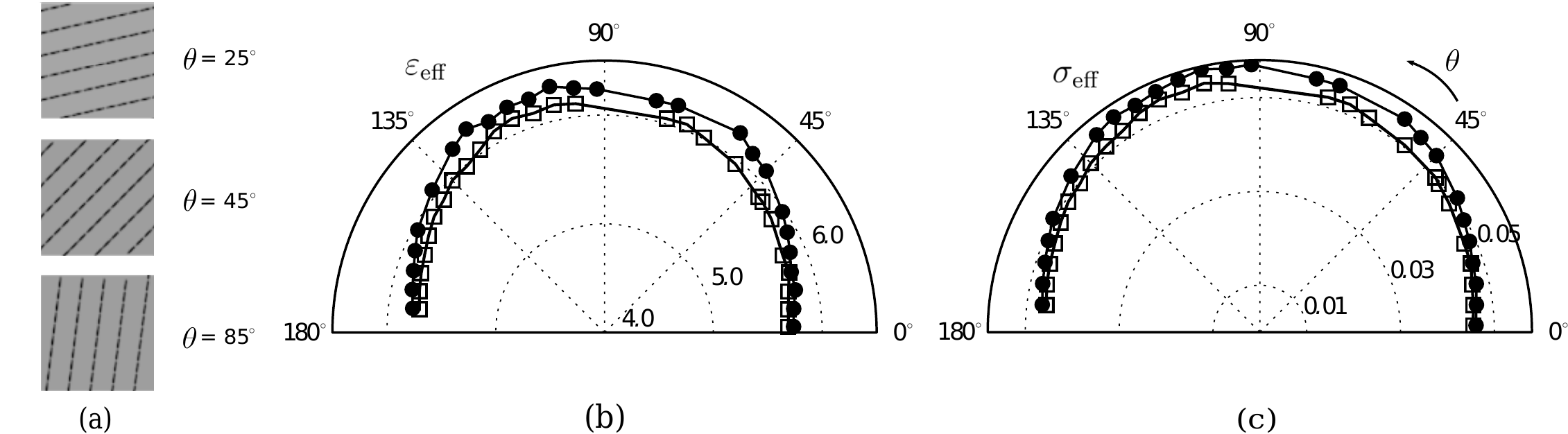}
\caption{Ellipse models. Percolated ($\bullet$) and not percolated ($\square$) models. The lines are guides for the eyes. (a) Simulation slice examples. (b)-(c) Polar plots for effective relative permittivity and effective conductivity, respectively. The variable $\theta$ is the slope angle of the oriented ellipses, which is coincident with the slope of the lines (see text for explanation).}
\label{polarPerm}
\end{figure}

Validation of the method is performed using an hexagonal array of discs, for which we have an analytic solution for the effective dielectric constant and the effective conductivity \cite{Perrins_1979} (see Fig. \ref{Fighexagonal}). Figures \ref{Fighexagonal} (a) and (b) correspond to inclusions of air ($\varepsilon_{\text{i}}=0$, $\sigma_{\text{i}}=0$) embedded in a matrix with $\varepsilon_{\text{e}}=16$ and $\sigma_{\text{e}}=0.043$ S/m. Figures \ref{Fighexagonal} (c) and (d) correspond to a model of trabecular bone tissue. The inclusions correspond to the trabeculae ($\varepsilon_{\text{i}}=20.584$, $\sigma_{\text{i}}=0.364$ S/m), and the matrix is the bone marrow ($\varepsilon_{\text{e}}=5.485$, $\sigma_{\text{e}}=0.043$ S/m). These validation technique were also used by Bonifasi-Lista and Charkaev \cite{bonifasi2009electrical}. The fraction occupied by inclusions is varied by changing the disc radius. Good agreement is observed 
between the prediction made by the minimisation of the response function difference with an 
effective homogeneous medium and the analytical $\varepsilon_{\text{eff}}$ and $\sigma_{\text{eff}}$.\\
Disc shape inclusions that are randomly located are also tested (see Fig. \ref{elipsescotas} (a)) with values that are similar to those used in reference \cite{karkkainen2000}. An excellent agreement between the results presented here and in reference \cite{karkkainen2000} is observed. Different circle radii are tested; it can be seen that the permittivity and conductivity do not depend on the radius but instead depend on the occupied fraction ($\phi_{\text{i}}$).\\
It is known that the effective dielectric properties are determined not only by the fraction occupied by the inclusions but also by their shape \cite{mejdoubi2006}. Ellipse-shaped inclusion media are tested with effective formulas obtained in reference \cite{mejdoubi2006} (see \ref{apendix1}). The orientation $\gamma$ of the ellipses is a random with normal distribution ($\gamma \sim N(\hat{\gamma},\sigma^{2}=100)$). Three mean values of orientation are considered here, specifically $\hat{\gamma}=$ 0$^{o}$, 45$^{o}$ and 90$^{o}$. Overlapping of inclusions is allowed. Figure \ref{elipsescotas} (b) shows a comparison between ellipses ($\hat{\gamma}=$ 45$^{o}$) with discs and the theoretical bounds of \ref{apendix1} (Eqs. \ref{Wienermax}-\ref{HSmin}). In Figs. \ref{elipsescotas} (c) and (d) a comparison between ellipse-shaped inclusion media and theoretical bounds (Eq.\ref{eqMejdoubi}) is shown. Although these models 
were fitted for $\phi_{\text{i}} < 0.1$, the symmetric Bruggeman approach (using $\beta_{\text{SBG}}$), 
they show relatively good agreement for an occupied fraction of up to 0.2 (see dashed-dot and crosses marked in Figs.\ref{elipsescotas} (c) and (d)).\\
Figure \ref{polarPerm} (a) shows an example of a maximally anisotropic model, which comprises in ellipses oriented and arranged along parallel lines. The slope $\theta$ of the lines is varied (see Fig. \ref{polarPerm} (a)). The effective dielectric properties are calculated as a function of $\theta$. The source is always polarised, called $E_{y}$, and $\phi_{\text{i}}\approx$ 0.054. Two types of models are tested: (i) allowing the overlap of ellipses, and (ii) not allowing overlaps.  These models are simulated to study whether percolation dominates the effective dielectric behavior. It is noted that inclusions can form connected sets when overlapping each other. In this specific example, a percolated system is obtained when there are connected paths from one border to the other of the simulation box. Note that, at the simulation frequency of this work (950 
MHz), percolated (when overlapping is allowed) and non-percolated systems have similar behavior. The M-B model using $\beta_{\
text{SBG}}$ gives a depolarisation factor of $A=$ 0.901 for $E_{y}\perp MIL_{\text{max}}$ and $A=$ 0.120 for $E_{y}\shortparallel \text{MIL}_{\text{max}}$, which results in $\varepsilon_{\text{eff}} = $ 5.729, 6.120 and $\sigma_{\text{eff}} = $ 0.046, 0.054 S/m, respectively. The values obtained by numerical simulations are $\varepsilon_{\text{eff}} = $ 5.689, 6.120 and $\sigma_{\text{eff}} = $ 0.045, 0.053 S/m. We see that 
the agreement between the simulations and effective models is relatively good.
\subsection{Simulation on realistic models}
\begin{figure}
\centering
\includegraphics[width=0.7\textwidth]{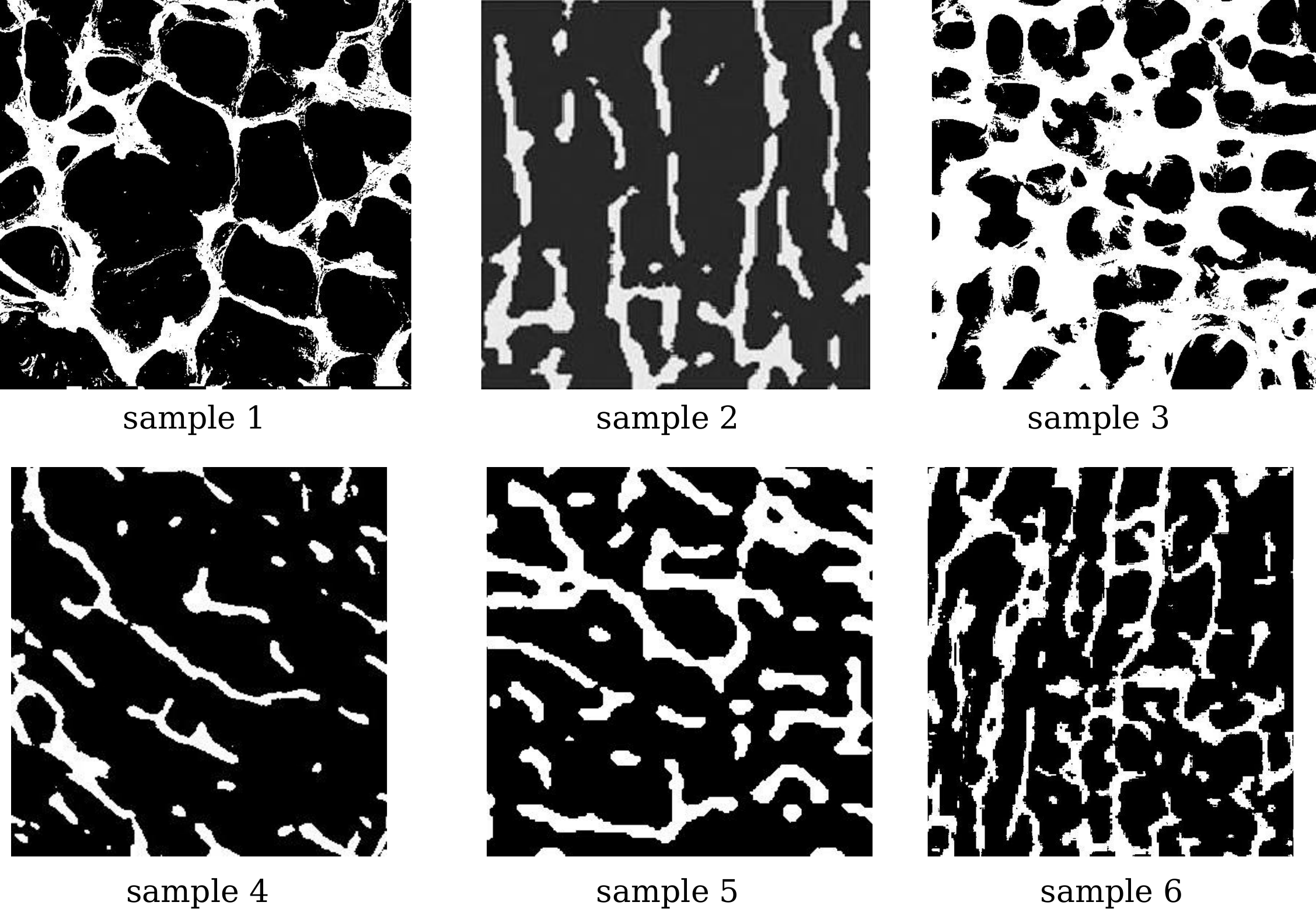}
\caption{Trabecular bone tissue samples used in the simulation of realistic models obtained from \cite{hansmalab,hosokawa2009numerical}.}
\label{samples}
\end{figure}

\begin{figure}%[h]
\centering
\includegraphics[width=1\textwidth]{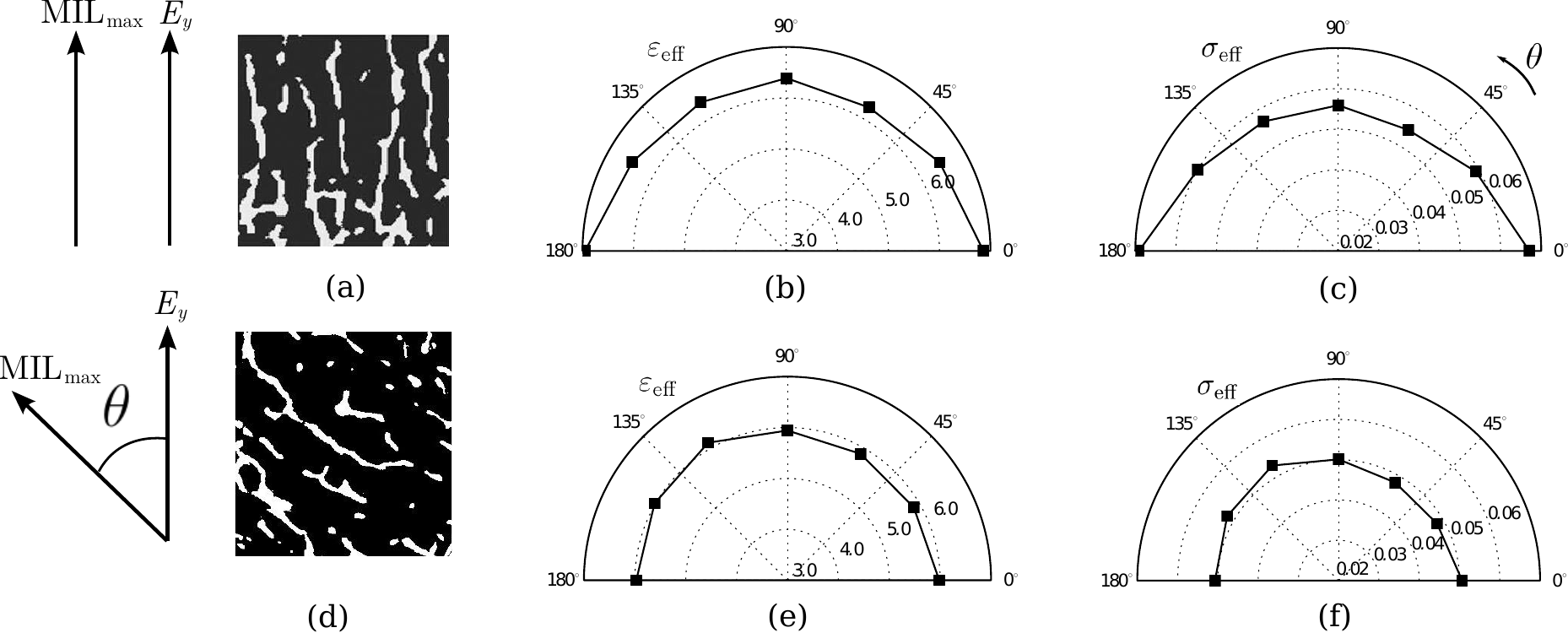}
\caption{Effective dielectric properties of sample 2 (a-c) and sample 4 (d-f) as a function of the angle ($\theta$). Parts (a) and (d) show the samples and the main direction. Parts (b) and (e) show the effective relative permittivity. Parts (c) and (f) show the effective conductivity. The lines are visual guides.}
\label{muestrasrotando}
\end{figure}

Figure \ref{samples} shows the samples that were used to build the realistic models. The models are simulated as a two-phase medium in which the trabeculae are the inclusions and the background is the bone marrow. The images are rotated at an angle of $\theta$ (at 30 degree steps), and the simulation is conducted at each orientation (the same method, as explained in subsection \ref{NumModel}, is used to calculate the effective dielectric properties). The applied field is always $y$ polarised.\\
Table \ref{Table I} summarises the micro-structure parameters of each sample (calculated as described in subsection \ref{MicroParam}). There is only one sample from a young individual (sample 3). Sample 1 and 3 have the lower DA values. \\
\begin{table}
\centering
\caption{Micro-structure properties of the trabecular bone tissues of Fig. \ref{samples}.}
\label{Table I}
\begin{tabular}{cccccc}
\toprule
Sample & Porosity & BV/TV & $\eta$(MIL$_{\text{max}}$) & Trab. Thick. [mm]$\pm$std & DA\\
\hline
1 & 0.75 & 0.25 & 0$^{o}$ & 0.11$\pm$0.16 & 1.08\\ 
2 & 0.79 & 0.21 & 90$^{o}$ & 0.18$\pm$0.13 & 2.23\\ 
3 & 0.46 & 0.54 & 0$^{o}$ & 0.23$\pm$0.32 & 1.14\\  
4 & 0.88 & 0.12 & 130$^{o}$ & 0.13$\pm$0.08 & 1.53\\ 
5 & 0.74 & 0.26 & 90$^{o}$ & 0.21$\pm$0.19 & 1.40 \\ 
6 & 0.74 & 0.26 & 90$^{o}$ & 0.13$\pm$0.14 & 1.42\\
\bottomrule
\end{tabular}
\end{table}
Table \ref{Table II} shows the effective dielectric properties of the samples described in Table \ref{Table I}. $\theta_{\text{max}}$ and $\theta_{\text{min}}$ are the angles for which the dielectric properties are maximum ($\varepsilon_{\text{eff-max}}$ and $\sigma_{\text{eff-max}}$) and minimum ($\varepsilon_{\text{eff-min}}$ and $\sigma_{\text{eff-min}}$), respectively. It is important to notice that $\varepsilon_{\text{eff}}$ and $\sigma_{\text{eff}}$ are maximums when the applied field is totally aligned with the main direction of the sample ($\eta$(MIL$_{\text{max}}$)). In this table the ratios $\varepsilon_{\text{eff-max}}/ \varepsilon_{\text{eff-min}}$ and $\sigma_{\text{eff-max}}/ \sigma_{\text{eff-min}}$ are also shown. It is observed that the latter always gives larger values than the former. See, for example: sample 2; it has the maximum degree of anisotropy (2.23) and the maximum conductivity ratio (1.28), while the 
permittivity ratio is 1.12.\\
Figure \ref{muestrasrotando} shows typical simulations of sample 2 (Fig. \ref{muestrasrotando} (a)) and 4 (Fig. \ref{muestrasrotando} (b)). Polar plots show effective relative permittivity and conductivity as a function of $\theta$. Parts (b) and (c) correspond to sample 2 and (e) and (f) correspond to sample 4. It should be noted that the main direction can be detected with both the permittivity and conductivity.
\begin{table}
\centering
\caption{Effective dielectric properties obtained by the simulation of realistic models.}
\label{Table II}
\begin{tabular}{ccccccccc}
\toprule
Sample & $\theta_{\text{max}}$ & $\theta_{\text{min}}$  & $\varepsilon_{\text{eff-max}}$  & $\varepsilon_{\text{eff-min}}$ & $\dfrac{\varepsilon_{\text{eff-max}}}{\varepsilon_{\text{eff-min}}}$ & $\sigma_{\text{eff-max}}$ & $\sigma_{\text{eff-min}}$& $\dfrac{\sigma_{\text{eff-max}}}{\sigma_{\text{eff-min}}}$\\
\hline
1 & 90$^{o}$ & 30$^{o}$ & 7.503 & 7.096 & 1.06 & 0.077 & 0.0706 & 1.09\\ 
2 & 0$^{o}$ & 90$^{o}$ & 6.950  & 6.247 & 1.12 & 0.069 & 0.054 & 1.28\\ 
3 & 90$^{o}$ & 0$^{o}$ & 11.446 & 10.345 & 1.11 & 0.154 & 0.149 & 1.18\\  
4 & 120$^{o}$ & 30$^{o}$ & 6.119 & 5.864 & 1.04 & 0.053 & 0.048 & 1.10\\ 
5 & 0$^{o}$ & 60$^{o}$ & 7.312 & 6.798 & 1.07 & 0.075 & 0.063 & 1.19 \\ 
6 & 0$^{o}$ & 90$^{o}$  & 7.189 & 6.782 & 1.06 & 0.071 & 0.063 & 1.13\\
\bottomrule
\end{tabular}
\end{table}
\subsection{Effective models}
In this subsection, comparison between the effective models and realistic models are made. It was shown in \cite{mejdoubi2006} that the depolarisation factor (a scalar in 2D) for media with elliptical inclusions depends on the aspect ratio ($a/b$) and orientation. As shown in Fig. \ref{elipsescotas}, effective M-B models (using $\beta_{\text{SBG}}$) have a relatively good agreement when compared with simulation data for $\phi_{\text{i}}<0.2$. In general, this result was observed for several simulations that are not shown here. Table \ref{Table III} shows some results of the M-B models (using $\beta_{\text{SBG}}$) as a function of $\phi_{\text{i}}$. We replace $\phi_{\text{i}}$ by the $BV / TV$ parameter of each sample. Three aspect ratios are considered: $\frac{a}{b}=$1.0,1.5 and 3.0. Note that, for high values of DA, although the absolute values of effective dielectric properties ​​agree only in part, the 
correlation with the relative values of $\frac{a}{b}=$1.5 is good (values highlighted in boldface in Table \ref{Table III})
.
\begin{table}

\centering
\caption{Effective M-B models using $\beta_{\text{SBG}}$.}
\label{Table III}

\begin{tabular}{lccccccc} 
\toprule
 & & \multicolumn{5}{c}{$\phi_{\text{i}}\equiv BV / TV$}\\
\cline{3-7} 
$a/b$& &\small{0.12} & \small{0.21} & \small{0.25} & \small{0.26} & \small{0.54}\\
\cline{1-7} 
% \cline{3-7} 
\multirow{2}{*}{\small{1.0}} & $\varepsilon_{\text{eff}}$ & \small{6.688}&\small{7.103} &\small{7.342}& \small{7.565}&\small{10.703}\\\vspace{0.1in}
&$\sigma_{\text{eff}}$ &\small{0.065} &\small{0.066} & \small{0.071} & \small{0.073}&\small{0.124}\\

% \hline
\multirow{2}{*}{\small{1.5}} & $\frac{\varepsilon_{\text{eff-max}}}{\varepsilon_{\text{eff-min}}}$ &$\frac{6.471}{6.245}\approx$ \small{\textbf{1.04}}&$\frac{7.342}{6.910}\approx$ \textbf{\small{1.06}} & - & $\frac{7.873}{7.315}\approx$ \textbf{\small{1.07}}& -\\\vspace{0.1in}
& $\frac{\sigma_{\text{eff-max}}}{\sigma_{\text{eff-min}}}$ & $\frac{0.057}{0.053}\approx$ \small{\textbf{1.09}}& $\frac{0.071}{0.062} \approx$ \textbf{\small{1.14} }& - & $\frac{0.080}{0.068} \approx$ \textbf{\small{1.19}}& -\\
% \hline
\multirow{2}{*}{\small{3}} &$\frac{\varepsilon_{\text{eff-max}}}{\varepsilon_{\text{eff-min}}}$& $\frac{6.744}{6.129}\approx$ \small{1.10} & $\frac{7.844}{6.689}\approx$ \small{1.17}& - & $\frac{8.512}{7.028}\approx$ \small{1.21}& - \\
& $\frac{\sigma_{\text{eff-max}}}{\sigma_{\text{eff-min}}}$ & $\frac{0.064}{0.051}\approx$ \small{1.25}& $\frac{0.085}{0.058}\approx$ \small{1.47}& - & $\frac{0.098}{0.062}\approx$ \small{1.58}& - \\
\bottomrule
\end{tabular}
\end{table}

\section{Discussion}
\label{DISCUSS}
 Microwave tomography methods are developed in a wide frequency range. Among others, the experimental setup of Semenov et al. \cite{semenov2005microwave} uses 900 MHz and 2.05 GHz, Meaney et al.\cite{meaney2012bone} use 900, 1100 and 1300 MHz. The simulations of Liu et al. \cite{liu2002active} were performed at 800 MHz and 6 GHz, and the simulation of Mojabi and Lo Vetri \cite{mojabi2011novel} where performed at 1000 MHz. The results in this paper are limited to a central frequency (950 MHz); this choice was motivated by the fact that the first microwave in vivo image of trabecular bone tissue was obtained at a value that was close to this frequency \cite{golnabi2011microwave}. Consequently, we assume constant values of permittivity and conductivity for each component of bone tissue (trabeculae and bone marrow).\\
Recent studies have begun to consider microwave imaging of bone tissue \textit{in vitro} \cite{semenov2005microwave,meaney2012bone} and \textit{in vivo} \cite{golnabi2011microwave}. In general, the current interest is to find correlations between dielectric properties and bone mineral density or bone volume. It is surprising that the relationship between anisotropy and dielectric properties has been studied relatively little. We will mainly focus this discussion on two aspects: first, we will present the correlation between $BV/TV$ and the effective dielectric properties, and, second, we will comment on the results that involve anisotropy.\\
A comparison between the dielectric properties and the bone volume, in Tables \ref{Table I} and \ref{Table II}, shows a positive correlation. This correlation implies that, the higher the $BV/TV$ the higher effective permittivity and conductivity. This result is very intuitive and is in accordance with two phase models: when the fraction of inclusion grows (with higher relative permittivity), the effective permittivity and conductivity of the medium also grow. Differences are found when these simulation results are compared with measurements \textit{in vivo}. In reference \cite{golnabi2011microwave}, the authors found that an affected calcaneus (with a lower bone volume) has a 23\% higher relative permittivity when compared with a normal bone. It is necessary to remark that this result is based on only one sample and is without statistic support. The same research group has recently published a paper on \textit{in vitro} measurements of trabecular porcine femur samples (six bone specimens in saline solution) 
\cite{meaney2012bone}. They studied the change in dielectric properties as a 
function of mineralisation. The degree of mineralisation was altered (not naturally) by acid treatment, and their results are in agreement with their previous work \cite{golnabi2011microwave}. These results also agree with \cite{irastorzatesis}, in which the degree of mineralisation was measured and a negative correlation between the bone mineral density and the relative permittivity was found (for six samples of human osteoporotic trabecular samples in saline solution). \\
Some simple calculation can be performed with the help of effective models. For example, using Eq.\ref{eqMejdoubi} for a total aligned electric field and bone orientation with $\phi_{\text{i}}=0.26$, $\varepsilon_{\text{i}}=20.584$, $\sigma_{\text{i}}=0.364$ and $\varepsilon_{\text{e}}=78.00$, $\sigma_{\text{e}}=1.39$ (similar values to 0.9\% saline solution) gives 61.76 and 1.08, for the permittivity and conductivity, respectively. For $\phi_{\text{i}}=$0.54, the results 42.79 and 0.749. These results offer a possible explanation of the relation between $BV / TS$ and the effective dielectric properties (at least for \textit{in vitro} measurements): samples that have a lower $BV / TV$ have more space for water or saline solution, which increases their permittivity and conductivity. Special care must be taken when interpreting \textit{in vivo} measurements. What takes the place of bone in a 
resorption process? Is it predominantly high water content connective tissue/cells/blood (with high dielectric properties), or is it bone 
marrow (with low dielectric properties)? As the first image \textit{in vivo} has shown\cite{golnabi2011microwave}, it seems that a loss of bone tissue increases both the permittivity and conductivity. Then, the first case above could be the more feasible process.\\
Regarding anisotropy, Table \ref{Table II} shows that the polarised electric field aligned with the main direction of the sample gives a higher relative permittivity and conductivity. Minimum values are found if the electric field and the main direction are orthogonal.  This result shows the potential application of microwave imaging as a predictor of trabecular orientation. To develop a clinical microwave system to detect these orthogonal components is, of course, a technical complication.\\
Table \ref{Table II} also shows that the degree of anisotropy correlates rather well with the permittivity ratio ($\varepsilon_{\text{eff-max}}/\varepsilon_{\text{eff-min}}$). The maximum value of DA (2.23) for sample 2 gives also a maximum ratio for the relative permittivity (1.12). In this case, the permittivity is 12\% higher when the electric field is aligned with the main direction. The conductivity ratio (the right-most column of Table \ref{Table II}) shows the strongest relationship with DA, for the case mentioned before, the conductivity is 28\% higher (when there is a total alignment). As a concluding remark on these results: the conductivity could be a better estimate for the DA.\\
Theoretical effective ellipses models better approximate the simulations when the occupied fraction is low (strictly speaking for $\phi_{\text{i}}<$ 0.1). Table \ref{Table III} continues the application to trabecular bone tissue. A relative good agreement can be seen for $BV / TV < $ 0.26 (see Table \ref{Table III} values in boldface). For example, samples 2, 4, 5 and 6 have values that are similar to those given for a model with ellipses with an aspect ratio $a/b=$ 1.5. These results show an interesting trend: ellipse models represent well the samples that have a high degree of anisotropy and a low occupied fraction. Moreover, when samples have a low DA and a low occupied fraction, circular inclusion models appear to be better.  This result is interesting for 
simulation purposes. In the case of trabecular bone (as a hierarchical material), a knowledge of the effective formulas avoids the need for excessive grid refinement. These methods are known as homogenisation techniques; Maxwell-Garnett and symmetric Bruggeman are classical 
approaches. The simulations presented here show that these approaches can represent the trabecular bone tissue relatively well.\\
 Even though the permittivity and conductivity are good predictors of the trabecular bone anisotropy and BV/TV for the samples simulated here, there remains a challenging question: which is the best site of the body to assess the bone quality by microwave imaging? A reasonable assumption is the heel \cite{golnabi2011microwave} (which is widely used for ultrasound in vivo measurements). The calcaneus is mostly composed of trabecular bone and, as measured by Hildebrand et al. \cite{hildebrand1999direct}, has the higher values of DA (if it is compared with the spine, femur head and iliac crest). In \cite{hildebrand1999direct} the authors reported values of DA and BV/TV of approximately 1.75 and 0.12, respectively, which are not far from the samples simulated here. In future studies, we plan to use the methodology presented here on calcaneus 
images.\\
As a final conclusion, we can say that samples with a high degree of anisotropy can give up to a 20\% difference between the aligned and orthogonal conductivity values. The relative permittivity shows a smaller variation. Semenov et al. \cite{semenov2005microwave} studied microwave-tomographic imaging of high dielectric-contrast objects (such as bone tissue and muscle or skin). Even when taking the bone as an isotropic tissue, the authors encountered several of the most difficult aspects of the reconstruction algorithms. Microwave imaging of bone tissue is still in its infancy, but according to the results presented here, anisotropy must be accounted for when measuring trabecular bone tissue.

\section*{Acknowledgments}
This work was supported by grants from Universidad Nacional de La Plata (Project 11/I153), Consejo Nacional de Investigaciones Cient\'ificas y
T\'ecnicas (PIP 1192) and Agencia Nacional de Promoci\'on Cient\'ifica y Tecnol\'ogica (PICT 00908) of Argentina. C.M.C. and F.V. are members of CONICET.

\section*{Conflict of interest statement}
No conflict of interest.

\section*{Ethical approval}
Not required.

\appendix

\section{Effective approaches}
\label{apendix1}
Wiener approach
\begin{equation}
\varepsilon_{\text{eff-max}}=\phi_{\text{i}}\varepsilon_{\text{i}}+(1-\phi_{\text{i}})\varepsilon_{\text{e}} \text{,}
\label{Wienermax}
%  \varepsilon_{ef,min}=\frac{\varepsilon_{i}\varepsilon_{e}}{(1-f)\varepsilon_{i}+f\varepsilon_{e}}
\end{equation}
\begin{equation}
\varepsilon_{\text{eff-min}}=\frac{\varepsilon_{\text{i}}\varepsilon_{\text{e}}}{(1-\phi_{\text{i}})\varepsilon_{\text{i}}+\phi_{\text{i}}\varepsilon_{\text{e}}} \text{.}
\label{Wienermin}
\end{equation}
Hashin-Strikman approach
\begin{equation}
\varepsilon_{\text{eff-max}}=\varepsilon_{\text{e}}+\frac{\phi_{\text{i}}}{\frac{1}{\varepsilon_{\text{i}}-\varepsilon_{\text{e}}}+\frac{1-\phi_{\text{i}}}{2\varepsilon_{\text{e}}}}\text{,}
\label{HSmax}
\end{equation}
\begin{equation}
\varepsilon_{\text{eff-min}}=\varepsilon_{\text{i}}+\frac{1-\phi_{\text{i}}}{\frac{1}{\varepsilon_{\text{e}}-\varepsilon_{\text{i}}}+\frac{\phi_{\text{i}}}{2\varepsilon_{\text{i}}}}  \text{.}
\label{HSmin}
\end{equation}
Mejdoubi-Brosseau approach
\begin{equation}
\varepsilon_{\text{eff}}=\varepsilon_{\text{e}}\cdot f\left(\frac{\varepsilon_{\text{i}}}{\varepsilon_{\text{e}}},\phi_{\text{i}},A \right)\approx\varepsilon_{\text{e}}\left(1+\alpha\phi_{\text{i}}+\beta\phi_{\text{i}}^{2}\right)\text{,}
\label{eqMejdoubiA}
\end{equation}
with $\alpha = 1 / [A+1 / (\varepsilon_{\text{i}} / \varepsilon_{\text{e}} -1)]$ and $\beta$ can take two values depending on the approach (sub indexes MG and SBG for Maxwell Garnett and symmetric Bruggeman, respectively):
\begin{center}
$\beta_{\text{MG}} =\dfrac{1}{A+\dfrac{2}{\dfrac{\varepsilon_{\text{i}}}{\varepsilon_{\text{e}}}-1}+\dfrac{1}{A}\left(\dfrac{1}{\dfrac{\varepsilon_{\text{i}}}{\varepsilon_{\text{e}}}-1}\right)^{2}}$\text{,}\\
\end{center}
or
\begin{center}
$\beta_{\text{SBG}} = \dfrac{\dfrac{\varepsilon_{\text{i}}}{\varepsilon_{\text{e}}}}{A^{2}\left(\dfrac{\varepsilon_{\text{i}}}{\varepsilon_{\text{e}}}-1\right)\left(1+\dfrac{1}{A\left(\dfrac{\varepsilon_{\text{i}}}{\varepsilon_{\text{e}}}-1\right)} \right)^{3}}$.
\end{center}

\bibliographystyle{elsarticle-num}
% \bibliography{references.bib}

\begin{thebibliography}{10}
\expandafter\ifx\csname url\endcsname\relax
  \def\url#1{\texttt{#1}}\fi
\expandafter\ifx\csname urlprefix\endcsname\relax\def\urlprefix{URL }\fi
\expandafter\ifx\csname href\endcsname\relax
  \def\href#1#2{#2} \def\path#1{#1}\fi

\bibitem{haidekker2011medical}
M.~Haidekker, G.~Dougherty, {Medical Imaging in the Diagnosis of Osteoporosis
  and Estimation of the Individual Bone Fracture Risk}, Springer, 2011.

\bibitem{kanis2008assessment}
J.~Kanis, W.~H. O. C. f. M.~B. Diseases, {Assessment of osteoporosis at the
  primary health care level}, WHO Collaborating Center for Metabolic Bone
  Diseases, University of Sheffield Medical School, 2008.

\bibitem{cherkaev2011characterization}
E.~Cherkaev, C.~Bonifasi-Lista, {Characterization of structure and properties
  of bone by spectral measure method}, Journal of Biomechanics 44~(2) (2011)
  345--351.

\bibitem{bonifasi2009electrical}
C.~Bonifasi-Lista, E.~Cherkaev, {Electrical impedance spectroscopy as a
  potential tool for recovering bone porosity}, Physics in medicine and biology
  54 (2009) 3063.

\bibitem{sierpowska2007effect}
J.~Sierpowska, M.~Lammi, M.~Hakulinen, J.~Jurvelin, R.~Lappalainen, J.~Toyras,
  {Effect of human trabecular bone composition on its electrical properties},
  Medical engineering \& physics 29~(8) (2007) 845--852.

\bibitem{sierpowska2007electrical}
J.~Sierpowska, {Electrical and Dielectric Characterization of Trabecular Bone
  Quality}, Ph.D. thesis, University of Kuopio, 2007.

\bibitem{sierpowska2005prediction}
J.~Sierpowska, M.~Hakulinen, J.~T{\"o}yr{\"a}s, J.~Day, H.~Weinans,
  J.~Jurvelin, R.~Lappalainen, {Prediction of mechanical properties of human
  trabecular bone by electrical measurements}, Physiological measurement 26
  (2005) S119.

\bibitem{sierpowska2003electrical}
J.~Sierpowska, J.~T{\"o}yr{\"a}s, M.~Hakulinen, S.~Saarakkala, J.~Jurvelin,
  R.~Lappalainen, {Electrical and dielectric properties of bovine trabecular
  bone relationships with mechanical properties and mineral density}, Physics
  in medicine and biology 48 (2003) 775.

\bibitem{sierpowska2006interrelationships}
J.~Sierpowska, M.~Hakulinen, J.~T{\"o}yr{\"a}s, J.~Day, H.~Weinans,
  I.~Kiviranta, J.~Jurvelin, R.~Lappalainen, {Interrelationships between
  electrical properties and micro-structure of human trabecular bone}, Physics
  in medicine and biology 51 (2006) 5289.

\bibitem{golnabi2011microwave}
A.~Golnabi, P.~Meaney, S.~Geimer, T.~Zhou, K.~Paulsen, {Microwave tomography
  for bone imaging}, in: {Biomedical Imaging: From Nano to Macro, 2011 IEEE
  International Symposium on}, IEEE, 2011, pp. 956--959.

\bibitem{meaney2012bone}
P.~Meaney, T.~Zhou, D.~Goodwin, A.~Golnabi, E.~Attardo, K.~Paulsen, {Bone
  Dielectric Property Variation as a Function of Mineralization at Microwave
  Frequencies}, International Journal of Biomedical Imaging 2012.

\bibitem{irastorzatesis}
R.~M. Irastorza,
  \href{http://sedici.unlp.edu.ar/?id=ARG-UNLP-TPG-0000001486}{{Identificaci{\'o}n
  de sistemas: aplicaci{\'o}n a la evaluaci{\'o}n in vitro de la calidad
  {\'o}sea en tejido trabecular humano}}, Ph.D. thesis, National University of
  La Plata, Argentina. (Sep. 2010).
\newline\urlprefix\url{http://sedici.unlp.edu.ar/?id=ARG-UNLP-TPG-0000001486}

\bibitem{peyman2011dielectric}
A.~Peyman, {Dielectric properties of tissues; variation with age and their
  relevance in exposure of children to electromagnetic fields; state of
  knowledge}, Progress in Biophysics and Molecular Biology.

\bibitem{Odgaard_book}
A.~Odgaard, {Bone Mechanics}, CRC Press LLC, 2001.

\bibitem{orwoll1999osteoporosis}
E.~Orwoll, {Osteoporosis in Men}, Wiley Online Library, 1999.

\bibitem{hosokawa2009numerical}
A.~Hosokawa, {Numerical analysis of variability in ultrasound propagation
  properties induced by trabecular micro-structure in cancellous bone},
  Ultrasonics, Ferroelectrics and Frequency Control, IEEE Transactions on
  56~(4) (2009) 738--747.

\bibitem{hosokawa2010effect}
A.~Hosokawa, {Effect of porosity distribution in the propagation direction on
  ultrasound waves through cancellous bone}, Ultrasonics, Ferroelectrics and
  Frequency Control, IEEE Transactions on 57~(6) (2010) 1320--1328.

\bibitem{saha1995comparison}
S.~Saha, P.~Williams, {Comparison of the electrical and dielectric behavior of
  wet human cortical and cancellous bone tissue from the distal tibia}, Journal
  of orthopaedic research 13~(4) (1995) 524--532.

\bibitem{teixeira2008time}
F.~Teixeira, {Time-domain finite-difference and finite-element methods for
  Maxwell equations in complex media}, Antennas and Propagation, IEEE
  Transactions on 56~(8) (2008) 2150--2166.

\bibitem{karkkainen2000}
K.~K{\"a}rkk{\"a}inen, A.~H. Sihvola, K.~I. Nikoskinen, {Effective permittivity
  of mixtures: numerical validation by the FDTD method}, IEEE Transactions on
  Geoscience and Remote Sensing 38~(3) (2000) 1303--1308.

\bibitem{mejdoubi2006}
A.~Mejdoubi, C.~Brosseau, {Finite-element simulation of the depolarization
  factor of arbitrarily shaped inclusions}, Physical Review E 74~(3) (2006)
  031405.

\bibitem{oskooi2010}
A.~Oskooi, D.~Roundy, M.~Ibanescu, P.~Bermel, J.~Joannopoulos, S.~Johnson,
  {MEEP: A flexible free-software package for electromagnetic simulations by
  the FDTD method}, Computer Physics Communications 181~(3) (2010) 687--702.

\bibitem{taflove2005}
S.~Hagness, A.~Taflove, {Computational electrodynamics: The finite-difference
  time-domain method}, 3rd Edition, Artech House, Inc., 2005.

\bibitem{webdielectric}
\href{http://niremf.ifac.cnr.it/tissprop}{{Dielectric properties of body
  tissues}}.
\newline\urlprefix\url{http://niremf.ifac.cnr.it/tissprop}

\bibitem{gabriel1996dielectric}
S.~Gabriel, R.~Lau, C.~Gabriel, {The dielectric properties of biological
  tissues: III. Parametric models for the dielectric spectrum of tissues},
  Physics in medicine and biology 41 (1996) 2271.

\bibitem{scipy}
\href{http://www.scipy.org}{{Scientific Tools for Python}}.
\newline\urlprefix\url{http://www.scipy.org}

\bibitem{hansmalab}
\href{http://hansmalab.physics.ucsb.edu}{{Paul Hansma Research Group}}.
\newline\urlprefix\url{http://hansmalab.physics.ucsb.edu}

\bibitem{scikits}
\href{http://scikits-image.org}{{Scikits-image, image processing in python}}.
\newline\urlprefix\url{http://scikits-image.org}

\bibitem{Perrins_1979}
W.~T. Perrins, D.~R. McKenzie, R.~C. McPhedran, {Transport properties regular
  of arrays of cylinder}, Proc. R. Soc. Lond. A 369 (1979) 207--225.

\bibitem{semenov2005microwave}
S.~Semenov, A.~Bulyshev, A.~Abubakar, V.~Posukh, Y.~Sizov, A.~Souvorov,
  P.~van~den Berg, T.~Williams, {Microwave-tomographic imaging of the high
  dielectric-contrast objects using different image-reconstruction approaches},
  Microwave Theory and Techniques, IEEE Transactions on 53~(7) (2005)
  2284--2294.

\bibitem{liu2002active}
Q.~Liu, Z.~Zhang, T.~Wang, J.~Bryan, G.~Ybarra, L.~Nolte, W.~Joines, {Active
  microwave imaging. I. 2-D forward and inverse scattering methods}, Microwave
  Theory and Techniques, IEEE Transactions on 50~(1) (2002) 123--133.

\bibitem{mojabi2011novel}
P.~Mojabi, J.~LoVetri, {A novel microwave tomography system using a rotatable
  conductive enclosure}, Antennas and Propagation, IEEE Transactions on 59~(5)
  (2011) 1597--1605.

\bibitem{hildebrand1999direct}
T.~Hildebrand, A.~Laib, R.~M{\"u}ller, J.~Dequeker, P.~R{\"u}egsegger, {Direct
  Three-Dimensional Morphometric Analysis of Human Cancellous Bone:
  Microstructural Data from Spine, Femur, Iliac Crest, and Calcaneus}, Journal
  of Bone and Mineral Research 14~(7) (1999) 1167--1174.

\end{thebibliography}

\end{document}